\documentstyle[11pt] {article}

\title{The Stability of Lorentzian Space-Time}
\author{
Asher Yahalom$^{a,b}$\\
$^a$ Institute of Astronomy, University of Cambridge,\\
Madingley Road, Cambridge CB3 0HA, United Kingdom\\
$^b$ College of Judea and Samaria, Ariel 44284, Israel\\
e-mail: asya@ast.camb.ac.uk}

\begin{document}
\maketitle

\newcommand{\beq} {\begin{equation}}
\newcommand{\enq} {\end{equation}}
\newcommand{\ber} {\begin {eqnarray}}
\newcommand{\enr} {\end {eqnarray}}
\newcommand{\eq} {equation}
\newcommand{\eqs} {equations }
\newcommand{\mn}  {{\mu \nu}}
\newcommand{\sn}  {{\sigma \nu}}
\newcommand{\rhm}  {{\rho \mu}}
\newcommand{\sr}  {{\sigma \rho}}
\newcommand{\bh}  {{\bar h}}
\newcommand {\er}[1] {equation (\ref{#1}) }

\begin {abstract}
It is stated in many text books that the any metric appearing in
general relativity should be locally Lorentzian i.e. of the type
$\eta_\mn = \ {\rm diag } \ (1,-1,-1,-1)$ this is usually
presented as an independent axiom of the theory, which can not be
deduced from other assumptions. In this work we show that the
above assertion is a consequence of a standard stability analysis
of the Einstein \eqs and need not be assumed.

\bigskip
PACS: 03.30.+p, 04.20.Cv

\bigskip
keywords: General Relativity; Stability of Solutions;

\end {abstract}

\section {Introduction}

It is well known that our daily space-time is approximately
of Lorentz \\ (Minkowski) type that is,
it possess the metric $\eta_{\mn} = \ {\rm diag } \ (1,-1,-1,-1)$.
The above statement is taken as one of the central assumptions of the theory
of special relativity.

Further more it is assumed in the general theory of relativity that any space-time
is locally of the type $\eta_{\mn} = \ {\rm diag } \ (1,-1,-1,-1)$, although
it can not be presented so globally due to the effect of matter. This is
a part of the demands dictated by the well known equivalence principle.
The above principle is taken to be one of the assumptions of general relativity
in addition to the Einstein equations:
\beq
G_\mn = -\frac{8 \pi G}{c^4} T_\mn
\label{ein}
\enq
in which $G_\mn$ is the Einstein tensor, $T_\mn$ is the stress-energy tensor,
$G$ is the gravitational constant and $c$ is the velocity of light.

In what follows we will show that such assumption is not necessary,
(contrary to what is argued in so many text books, see for example
 \cite{MTW}) rather we will argue that this metric is the only possible stable solution
to the Einstein \er{ein} in vacuum, that is for the case $T_\mn = 0$.
And thus reduce the number of assumptions needed to obtain the
celebrated results of general relativity. By making the theory more compact
we enhance its predictive strength.

Eddington \cite[page 25]{Edd} has considered the possibility that the
universe contains different domains in which some domains are locally
Lorentzian and others have some other local metric of the type
 $\eta_{\mn} = \ {\rm diag } \ (-1,-1,-1,-1)$ or the type
$\eta_{\mn} = \ {\rm diag } \ (+1,+1,-1,-1)$.  For the first
case he concluded that the transition will not be possible since
one will have to go through a static universe with a metric
$\eta_{\mn} = \ {\rm diag } \ (0,-1,-1,-1)$\footnote{Prof. Lynden Bell
has noticed that there may be another way going through the metric
$\eta_{\mn} = \ {\rm diag } \ (\infty,-1,-1,-1)$, the author thanks
him for his remark.}. Going to the domain in which $\eta_{\mn} = \ {\rm diag } \ (+1,+1,-1,-1)$
means that one will have to pass through $\eta_{\mn} = \ {\rm diag } \ (+1,0,-1,-1)$
in which space becomes two dimensional\footnote{Again there may be another way going through the metric
$\eta_{\mn} = \ {\rm diag } \ (+1,\infty,-1,-1)$.}. The stability of those
domains was not discussed by Eddington.

Greensite \cite{Greensite1} and Carlini \& Greensite \cite{Greensite2,Greensite4}
have studied the metric $\eta_{\mn} = \ {\rm diag } \ (e^{i\theta},-1,-1,-1)$
in which $\theta$ the "wick angle" was treated as a quantum field dynamical variable. They have
shown that the real part of the quantum field effective potential is minimized for the Lorentzian
metric $\theta=0$ and for the same case the imaginary part of the quantum field effective potential is
stationary. Further more they have calculated the fluctuations around this minimal value and
have shown them to be of the order $(\frac{l_p}{R})^3$ in which $l_p$ is the Planck length and $R$ is the
scale of the universe. Elizalde \& collaborators \cite{Greensite6} have shown that the same arguments
apply to a five dimensional Kaluza-Klein universe of the type $R^4 \times T^1$.

Itin \& Hehl \cite{Itin} have deduced that space time must have a Lorentzian metric in order
to support classical electric/magnetic reciprocity.

H. van Dam \& Y. Jack Ng \cite{vanDam} have argued that in the absence of a Lorentzian metric
one can not obtain an appropriate finite representation of space-time and hence the various quantum
wave equations can not be written.

What is common to the above approaches is that
additional theoretical structures \& assumptions are needed in order to justify what appears
to be a fundamental property of space-time. In this paper we claim otherwise.
We will show that General relativistic equations and some "old fashioned" stability analysis
will lead to a unique choice of the Lorentzian metric being the only one which is stable.

The plan of this paper is as follows: in the first section we describe the possible
constant metrics which are not equivalent to each other by trivial
manipulations. The second section will be devoted to the
 review the classical linearized
theory of gravity which we adapt to our more general case of a non
Lorentzian background metric. In the third section we use the linearized
theory to study the stability of the possible constant metrics and thus
divide them into two classes: stable and unstable.
The last section will discuss some possible implications of our
results.

\section{Possible Constant Metrics}

In this section we study what are the possible constant
metrics available in the general theory of relativity
which are not equivalent to one another by a trivial transformation,
that amounts to a simple change of coordinates.

Let us thus study the four-dimensional interval:
\beq
d\tau^2 = \eta_\mn dx^{\mu} dx^{\nu}
\label{inter}
\enq
in what follows Greek letters take the traditional values of $1-4$,
and summation convention is assumed. $\eta_\mn$ is any real constant
matrix.

Since $\eta_\mn$ is symmetric we can diagonalize it using a unitary
transformation in which both the transformation matrix and the eigenvalues
obtained are real. Thus without loss of generality we can assume that in
a proper coordinate basis:
\beq
\eta = {\rm diag} \ (\lambda_1,\lambda_2,\lambda_3,\lambda_4).
\label{diag}
\enq

Next, by changing the units of the coordinates, we can always
obtain:
\beq
\eta = {\rm diag} \ (\pm 1,\pm 1,\pm 1,\pm 1)
\label{diag-norm}
\enq
notice that a zero eigen-value is not possible due to our assumption
that the space is four dimensional.

We conclude that the metrics $\eta$ given in \er{diag-norm}
 are the most general constant metrics possible.
In what follows we will study the stability of those solutions.

\section{Stability of Constant Metrics}

To study the stability of the metric $\eta_\mn$ given in \er{diag-norm}
we make an arbitrary small perturbation of the constant metric and obtain
the perturbed metric $g_\mn$.
\beq
g_\mn = \eta_\mn + h_\mn.
\label{perturb}
\enq
The evolution of the metric $g$ is studied under the Einstein \er{ein}.
In what follows we adapt the well known linearized theory of gravity
given in many text books such as \cite{MTW}, to our more generalized
needs. In order to obtain the linearized equations the following quantities
should be linearized:
\begin{enumerate}
\item The connection coefficients deduced from the metric:
\beq
\Gamma^\mu_{\alpha \beta} \equiv \frac{1}{2} g^{\mn}
(g_{\alpha \nu, \beta} + g_{\beta \nu, \alpha} - g_{\alpha \beta, \nu})
\label{Gam}
\enq
in which $,$ stand for partial derivative, and $g^{\mn}$ is the inverse matrix of
$g_{\mn}$.
\item The Ricci tensor $R_\mn$ which is deduced from the connection coefficients:
\beq
R_\mn \equiv \Gamma^{\alpha}_{\mn, \alpha}  -  \Gamma^{\alpha}_{\mu \alpha, \nu}  +
 \Gamma^{\alpha}_{\beta \alpha}  \Gamma^{\beta}_{\mn}  -
 \Gamma^{\alpha}_{\beta \nu}  \Gamma^{\beta}_{\mu \alpha}
\label{Rdef}
\enq
\item The Einstein tensor $G_\mn$ which is deduced from the
Ricci tensor and the Curvature Scalar:
\beq
G_\mn \equiv R_\mn - \frac{1}{2} g_\mn R, \qquad R \equiv g^\mn R_\mn
\label{Gdef}
\enq
\end{enumerate}

\subsection{Notations}

First we introduce some notations: the inverse metric of $g_\mn$ is
 $g^\mn$ given by:
\beq
g^\mn = \eta^\mn - h^\mn
\label{inperturb}
\enq
in which $\eta^\mn$ is the inverse matrix of $\eta_\mn$,
by virtue of \er{diag-norm} it is also identical to it
i.e., $\eta^\mn=\eta_\mn$. An easy calculation will
 show that to first order in $h$ we obtain:
\beq
h^\mn =  \eta^\sn \eta^\rhm h_\sr.
\label{inh}
\enq
Further more we introduce the following notations:
\beq
h^{\mu}_{\sigma} = \eta^\mn h_\sn \qquad
h = h^{\mu}_{\mu} = \eta^\mn h_\mn.
\label{not}
\enq
Generally speaking we use the constant metric $\eta$ to raise and lower
indices.

\subsection{The Connection Coefficients}

Let us now calculate the linearized form of the affine connection
which is given by \er{Gam}. Inserting \er{perturb} and \er{inperturb} and keeping
only the first order terms in $h$ we obtain:
\beq
\Gamma^\mu_{\alpha \beta} = \frac{1}{2} \eta^{\mn}
(h_{\alpha \nu, \beta} + h_{\beta \nu, \alpha} -
 h_{\alpha \beta, \nu})
\label{linGam}
\enq
which can also be written using \er{not} as:
\beq
\Gamma^\mu_{\alpha \beta} = \frac{1}{2}
({{h_{\alpha}}^{\mu}}_{,\beta} + {{h_{\beta}}^{\mu}}_{,\alpha} -
 h_{\alpha \beta,}^{\ \ \ \mu})
\label{linGam2}
\enq

\subsection{The Ricci Tensor}

The Ricci Tensor given in \er{Rdef} can be written in a linearized form
using the result of \er{linGam2} and the notation defined in \er{not}
\beq
R_\mn = \frac{1}{2} ({h_{\mu}^{\ \alpha}}_{,\alpha \nu} + {h_{\nu}^{\ \alpha}}_{,\alpha \mu} -
h_{\mn, \alpha}^{\ \ \ \ \alpha} - h_{,\mn} ).
\label{linR}
\enq
The linearized Curvature Tensor can be calculated from \er{linR}
\beq
R = \eta^\mn  R_\mn = h^{\mu \alpha}_{\ \ ,\mu \alpha} - h_{,\alpha}^{\ \ \alpha}.
\label{linC}
\enq

\subsection{The Einstein Tensor}

Finally we obtain the linearized form of the Einstein Tensor which can be calculated from
\eqs  (\ref{Gdef},\ref{linR},\ref{linC}):
\beq
G_\mn = \frac{1}{2} ({{h_{\mu}}^{\alpha}}_{,\alpha \nu} + {{h_{\nu}}^{\alpha}}_{,\alpha \mu} -
{h_{\mn, \alpha}}^{\alpha} - h_{,\mn} ) -
 \frac{1}{2} \eta_\mn ({h_{\alpha \beta,}}^{\alpha \beta} - {h_{,\alpha}}^{\alpha}).
\label{linG}
\enq
In order to simplify the above notation the following quantity is defined:
\beq
\bar h_\mn = h_\mn -  \frac{1}{2} \eta_\mn h
\label{hbar}
\enq
Using this definition, \er{linG} can be written as:
\beq
2 G_\mn = -{\bh_{\mn,\alpha}}^{\ \ \ \ \alpha} -
\eta_\mn {\bh_{\alpha \beta,}}^{\ \ \ \alpha \beta} +
{{\bh_{\mu \alpha,}}^{\ \ \ \alpha}}{}_\nu +
{{\bh_{\nu \alpha,}}^{\ \ \ \alpha}}{}_\mu.
\label{linG2}
\enq

\subsection{Gauge Transformation}

The metric being a tensor can always be transformed to another coordinate system by
the transformation:
\beq
g'_\sr (x')  = g_\mn \frac{\partial x^\mu}{\partial x'^\sigma}
 \frac{\partial x^\nu}{\partial x'^\rho}
\label{gtr}
\enq
Let us introduce the transformation:
\beq
x^\mu = x'^\mu - \xi^\mu
\label{xidef}
\enq
in which $\xi$ is same order of magnitude as $h$. Further more let us define:
\beq
g'_\mn (x')= \eta_\mn + h'_\mn (x).
\label{perturbp}
\enq
Inserting \eqs (\ref{perturb},\ref{perturbp},\ref{xidef}) into \er{gtr}
we obtain to first order in $h$:
\beq
h'_\mn = h_\mn - \xi_{\mu,\nu} - \xi_{\nu,\mu} \qquad (\xi_{\mu} = \eta^\mn \xi_{\nu})
\label{gtrl}
\enq
this is denoted as the "gauge transformation", by changing the coordinates infinitesimally
we can always obtain a new $h'$ which is different from the old $h$ and is related to it
by \er{gtrl}. Our obvious choice of gauge will be one that simplifies \er{linG2}.
Let us calculate the expression $\bh'_{\mu \alpha,}{}^{\alpha}$:
\beq
\bh'_{\mu \alpha,}{}^{\alpha}={\bh_{\mu \alpha,}}{}^{\alpha} - {\xi_{\mu,\alpha}}{}^{\alpha}.
\label{gagfix1}
\enq
We can always choose $\xi$ such that:
\beq
{\xi_{\mu,\alpha}}{}^{\alpha} = {\bh_{\mu \alpha,}}{}^{\alpha}
\label{gagfix2}
\enq
in which the equation above is a second order equation for the $\xi_\mu$'s.
This does not "fix" the gauge since we can always introduce a new gauge:
$\xi_{\mu}^{1} = \xi_{\mu} + \xi_{\mu}^{0}$ in which:
${\xi^{0}_{\mu,\alpha}}^{\alpha} = 0$. Choosing the gauge according to \er{gagfix2}
we obtain:
\beq
{\bh'_{\mu \alpha,}}{}^{\alpha}=0
\label{gagfix3}
\enq
writing \er{linG2} in terms of $h'$ and dropping the prime we see that:
\beq
2 G_\mn = -{\bh_{\mn, \alpha}}{}^{\alpha}.
\label{linG3}
\enq

\section{Stability Analysis}

In the lack of matter Einstein \er{ein} becomes $G_\mn = 0$
that is through \er{linG3} we obtain the following \eqs
for $\bh_\mn$:
\beq
\bh_{\mn, \alpha}{}^{\alpha}=0.
\label{lineq1}
\enq
Next we introduce the Fourier decomposition of $\bh_\mn$:
\beq
\bh_\mn = \frac{1}{(2 \pi)^\frac{3}{2}}
\int^\infty_{-\infty} A_\mn (x_0, \vec k)
e^{i \vec k \cdot \vec x} d^3 k,  \quad \vec k  = (k^1,k^2,k^3),
 \quad \vec x  = (x^1,x^2,x^3)
\label{fourdecom}
\enq
 Introducing the decomposition \er{fourdecom} into \er{lineq1}
leads to:
\beq
\eta^{00} \partial_0^2 A_\mn - \eta^{ij} k_i k_j A_\mn = 0
\label{lineq2}
\enq
in which $i,j$ are integers between $1-3$. Choosing $\eta^{00}=1$ we see that
the only way to avoid exploding solutions is to choose
$\eta^{ij} = {\rm diag} \ (-1,-1,-1)$,
thus one stable metric would be:
\beq
 \eta^{(1)} =  {\rm diag} \ (1,-1,-1,-1)
\label{stsol1}
\enq
alternatively we can choose $\eta^{00}=-1$ in this case the only way to avoid
exploding solutions is to choose
$\eta^{ij} = {\rm diag} \ (1,1,1)$,
thus a second stable metric would be:
\beq
 \eta^{(2)} =  {\rm diag} \ (-1,1,1,1)
\label{stsol2}
\enq
that is $ \eta^{(1)} = -\eta^{(2)}$ .

In the case that the universe has a spatial cyclic topology in one or
more directions the Fourier integral in this direction can be replaced by a Fourier series
such that we only have $k_i$'s of the type:
\beq
k_i = \frac{2 \pi n_i}{L_i}
\label{kcyclic}
\enq
in which $n_i$ is an integer and $L_i$ is the dimension of the spatially cyclic universe
in the $i$ direction.

\section{Conclusions}

We conclude  from \eqs (\ref{stsol1},\ref{stsol2}) that the only constant
stable solution is of a Lorentz (Minkowski) type.

For other constant solutions
we expect instabilities for $k_i \rightarrow \infty$ where $i$ depends on the
unstable solution chosen. Thus the instabilities vary on very small length scale
of which $\lambda = \frac{2 \pi}{k} \rightarrow 0$, this length  can be the
smallest for which the general theory of relativity is applicable, perhaps
the planck scale $\lambda =l_p= 1.616 10^{-35} m$, in that case an unstable solution will
last for about $t=\frac{\lambda}{c} = 5.39 10^{-44} sec$.
However, in the presence of matter this may take longer.
This may explain why in QED an unstable
Euclidean metric is used such that $ \eta =  {\rm diag} \ (1,1,1,1)$,
this is referred to as "wick's rotation" \cite{Wein}.
\\
\\
\noindent
{\bf Acknowledgement}
\\
\\
\noindent
The author would like to thank Prof. Jacob Bekenstein \& Prof. Robert Englman for useful discussions.
The author would like to thank Prof. Donald Lynden-Bell for reading this manuscript and making useful suggestions.

\end{document}